\documentclass{article}
\usepackage{graphicx,epsfig,amssymb,amsmath, changepage,float, cite,appendix, epstopdf, bm, verbatim}
\usepackage[font=small,labelfont=bf]{caption}
\usepackage[hmargin=3cm,vmargin=3cm]{geometry}
\usepackage{pgfplots}
\usepackage{tikz}
\usepackage{color}
\usetikzlibrary{positioning}
\usetikzlibrary{decorations.markings}
\usetikzlibrary{calc}

\newcommand{\pd}[2]{\frac{\partial #1}{\partial #2}}

\newcommand{\TheTitle}{Radiation from structured-ring resonators}
\newcommand{\TheAuthors}{B. Maling\thanks{Department of Mathematics, Imperial College London, SW7 2AZ, U.K.},\phantom{s}O. Schnitzer\footnotemark[1]\phantom{s} \& R. V. Craster\footnotemark[1]  }

\numberwithin{equation}{section}
\numberwithin{figure}{section}
\numberwithin{table}{section}

\title{\TheTitle}
\author{\TheAuthors}

\nocite{*}

\begin{document}

\maketitle

\begin{abstract}
We investigate the scalar-wave resonances of systems composed of identical Neumann-type inclusions arranged periodically around a circular ring. Drawing on natural similarities with the undamped Rayleigh--Bloch waves supported by infinite linear arrays, we deduce asymptotically the exponentially small radiative damping in the limit where the ring radius is large relative to the periodicity. 
In our asymptotic approach, locally linear Rayleigh--Bloch waves that attenuate exponentially away from the ring are  matched to a ring-scale WKB-type wave field. The latter provides a descriptive physical picture of how the mode energy is transferred via tunnelling to a circular evanescent-to-propagating transition region a finite distance away from the ring, from where radiative grazing rays emanate to the far field. Excluding the zeroth-order standing-wave modes, the position of the transition circle bifurcates with respect to clockwise and anti-clockwise contributions, resulting in striking spiral wavefronts.
\end{abstract}

\section{Introduction}\label{sec:intro}

Sharp resonances associated with electromagnetic whispering gallery modes have led to their application in a wide variety of fields~\cite{vahala_03,whisper_rev}: resonators with circular or spherical boundaries are commonly employed as nano-scale sensors~\cite{vollmer_08,Soria_11}; filters~\cite{dixit_01}; components in lasers~\cite{hall_89,kudryashov_99,bykov_95,hodgson_05}; and cavities for sensitive experiments into non-linear optics~\cite{ilchenko_04}, opto-mechanical coupling~\cite{hofer_10}, and other effects. Further to this, they can be used to couple energy between optical fibres or waveguides, resulting in frequency-dependent filters, optical switches and logic gates, designating them as fundamental building blocks of integrated photonics~\cite{rabus_07}. There are however drawbacks in terms of materials available for manufacture, limitations due to surface roughness, and maximum attainable Q-factors~\cite{righini11a}. Given the breadth of application, there is clear motivation to consider alternative approaches to confining wave energy. 

A closely-related strategy for designing a resonator is to exploit the fact that, for suitable material contrast, modes are strongly confined within a straight waveguide of one medium embedded within another. Deforming such a structure into a closed ring, and provided that the radius of curvature is large compared with the operating wavelength, allows guided waves to precess around the ring with little radiation loss. This interpretation explains the excellent performance of dielectric ring resonators \cite{melloni01a} and optical ring waveguide resonators \cite{chin98a}, and detailed theory supporting this is provided by an extensive literature on curved waveguides \cite{Heiblum:75,snyder83a}.

As an alternative to using curved waveguides, we can draw upon the extensive literature on array-guided waves, and use this as the starting point for a structured-ring resonator. It is well known that linear arrays of inclusions support strongly confined Rayleigh--Bloch waves, and these have application in electromagnetism such as Yagi--Uda antennas~\cite{Hurd_54,Sengupta_59}, in edge waves for coastlines~\cite{Evans_93}, as spoof surface plasmons~\cite{pendry04a}, and in elasticity~\cite{every08a,colquitt14a}. They have also attracted mathematical attention in terms of uniqueness and existence issues \cite{bonnet94a,Linton_02} as well as modelling studies~\cite{porter99a,thompson08a}. The ubiquitous nature of Rayleigh-Bloch waves suggests that structured-ring resonators based on curved periodic arrays would be applicable to a wide variety of physical settings. Further motivation to investigate such systems comes from studies of highly-conducting disks periodically decorated with dielectric-filled grooves, investigated in the context of spoof surface plasmons\cite{pors_12,paloma}, which similarly support localised resonances.

\begin{figure}[t]
  \centering
  \setlength{\unitlength}{\textwidth}
\includegraphics[width=.9\linewidth]{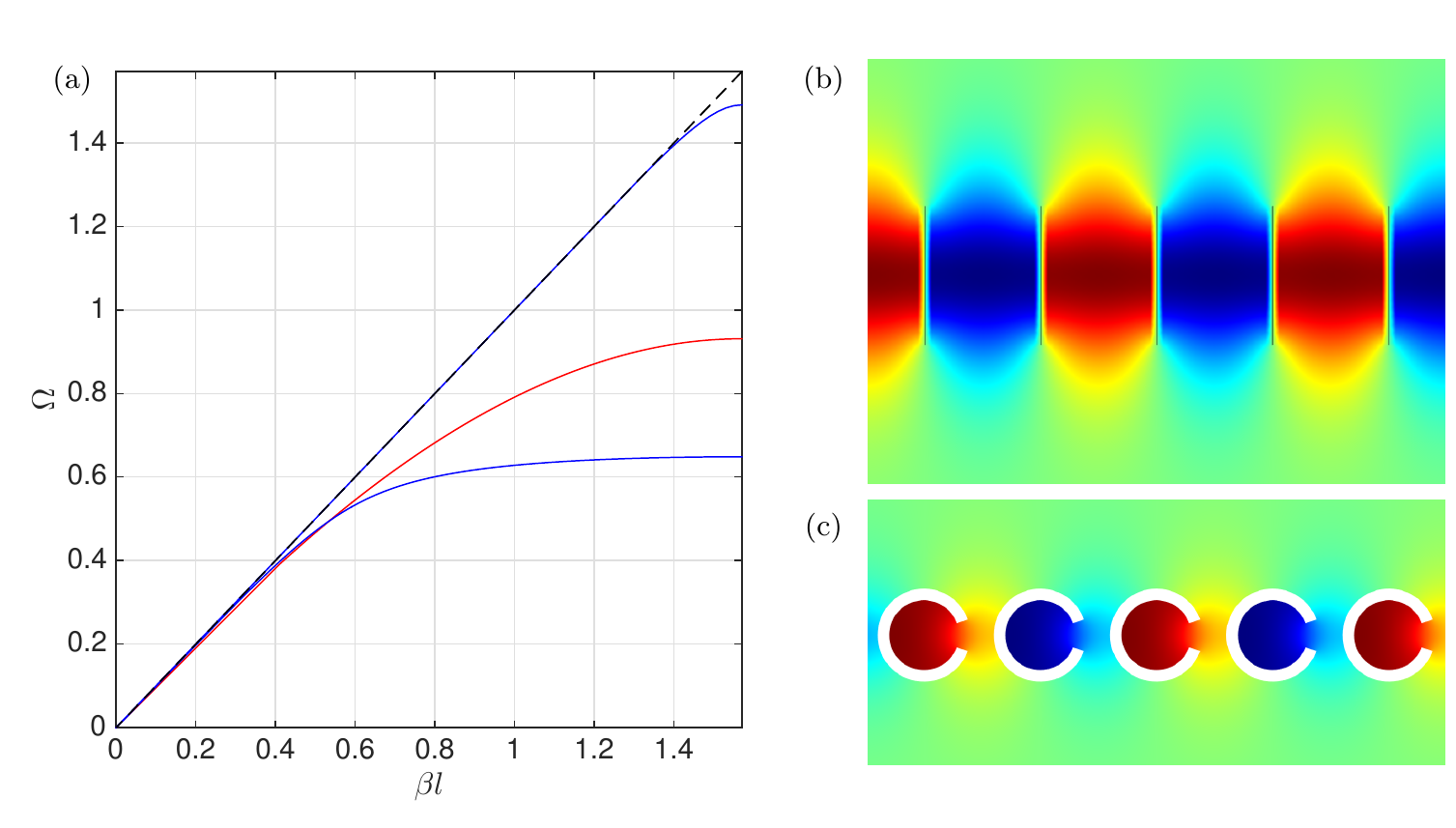}
\caption{Solutions of the Rayleigh--Bloch eigenvalue problem for two different linear arrays of homogeneous Neumann inclusions of period $2l$. The red dispersion curve in (a) is for an array of slit-like inclusions of height $1.2l$, and the blue curves are for C-shaped resonators formed of two concentric $320^{\circ}$ sectors with radii $0.8l$ and $0.6l$. Frames (b) and (c) show the standing waves at $\beta l=\pi/2$ for the dispersion branches ending at $\Omega\approx0.94$ and $\Omega\approx0.65$ respectively.}
\label{fig:RB}
\end{figure}

\begin{figure}[t]
  \centering
  \setlength{\unitlength}{\textwidth}
\includegraphics[width=\linewidth]{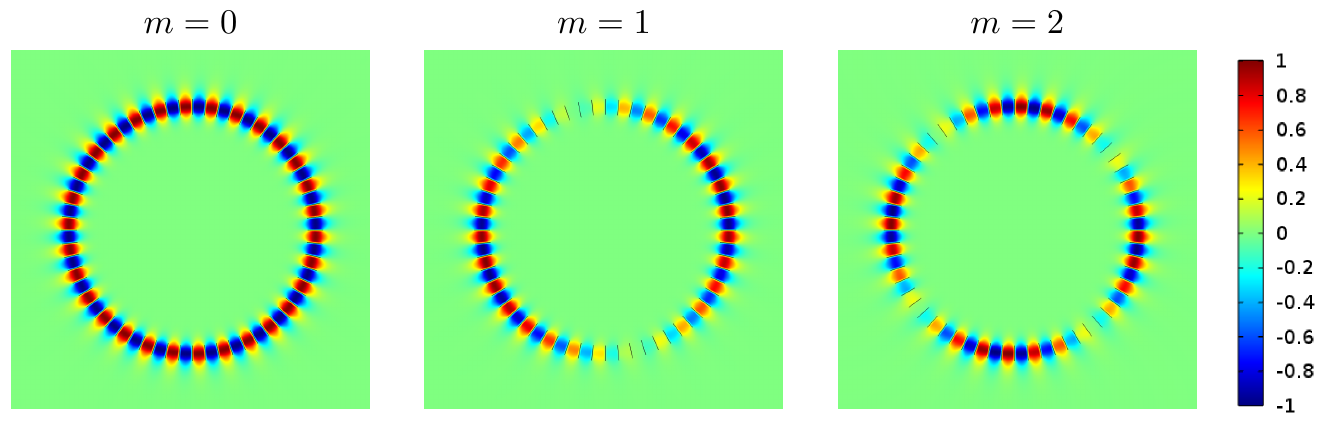}
\caption{Resonances of a unit ring of 60 slit-like homogeneous Neumann inclusions of length $1.2\pi/60$, for different values of the mode number $m$.}
\label{fig:wbm_trio}
\end{figure}

In this article we consider the scalar Helmholtz equation and arrays of Neumann inclusions for which Rayleigh--Bloch waves can exist \cite{linton02a}. For an infinite linear array, the frequencies of Bloch-periodic eigenfunctions lie on dispersion branches beneath the free-space light (sound) cone, as in figure \ref{fig:RB}(a), and hence the fields are strongly confined to the structure as they are unable to radiate energy; examples 
of such highly-confined Rayleigh--Bloch waves, computed numerically using finite element software \cite{comsol}, are illustrated in figure \ref{fig:RB}(b,c). We investigate resonances of circular rings created by the deformation of these linear arrays; three such resonances for a ring of slit-like inclusions are shown in figure \ref{fig:wbm_trio}. These geometries are a subset of those recently considered by two of the authors~\cite{maling_16_wbm}, in which a multiple scale asymptotic method was developed to investigate resonances with wide-angle modulation. Our aim here is to investigate analytically how the radiation damping, characterised by the Q-factor, of these resonances depends on the geometry and number of inclusions in the ring.

Asymptotic analysis provides a natural framework within which to analyse the radiation loss. 
We consider the limit where the number of inclusions $N$ is large and define $\epsilon=\pi/N\ll 1$, with the interpretation of $\epsilon$ being half of the angular period of the array, which will later be used as the asymptotic small parameter. We expect to find resonant modes that, in the vicinity of the structured ring, and to a leading-order approximation, coincide with the undamped Rayleigh--Bloch modes supported by the corresponding infinite linear-array configuration. Since Rayleigh--Bloch modes typically occur at wavelengths comparable to the array periodicity, and since the latter periodicity is assumed small compared with the ring radius, on the scale of the ring radius we are necessarily in the ray-optics regime of high-frequency wave propagation. Accordingly,  the problem separates into an inner region local to the ring that is then matched to an outer region where a solution in the form of a Wentzel$-$Kramers\(-\)Brillouin (WKB) ansatz is sought; the WKB method is well-known in the asymptotic and physics communities \cite{bender,hinch}, and specifically has been utilised in the related problem of a bent waveguide \cite{gridin03b}. The application of the method here is somewhat nonstandard, with the short-scale cyclic quantisation dictated by Bloch's theorem leading to unfamiliar terms and a bifurcation --- captured with a strained-coordinates ansatz --- of the `turning-point' problem with respect to contributions to the wave field propagating in the clockwise and anti-clockwise directions. 

The paper is structured as follows. 
In section \ref{sec:formulation} we formulate the eigenvalue problem governing the resonant modes of a structured ring, with the main goal set out in section \ref{sec:Qfactor} to analyse the exponentially small radiation damping. A detailed asymptotic analysis is carried out in section \ref{sec:analysis}, and then employed in sections \ref{sec:damping} and \ref{sec:pattern}, respectively, towards deriving explicit expressions for the radiation loss and field, the latter explaining spiral wavefields seen in numerical simulations. In section \ref{sec:delta}, we briefly discuss an intermediate-asymptotics regime in which the radiation damping is algebraically small and our asymptotic theory breaks down, and finally we draw together concluding remarks in section \ref{sec:conclude}.

\section{Formulation}\label{sec:formulation}
\begin{figure}[t]
  \centering
  \setlength{\unitlength}{\textwidth}
\includegraphics[width=\linewidth]{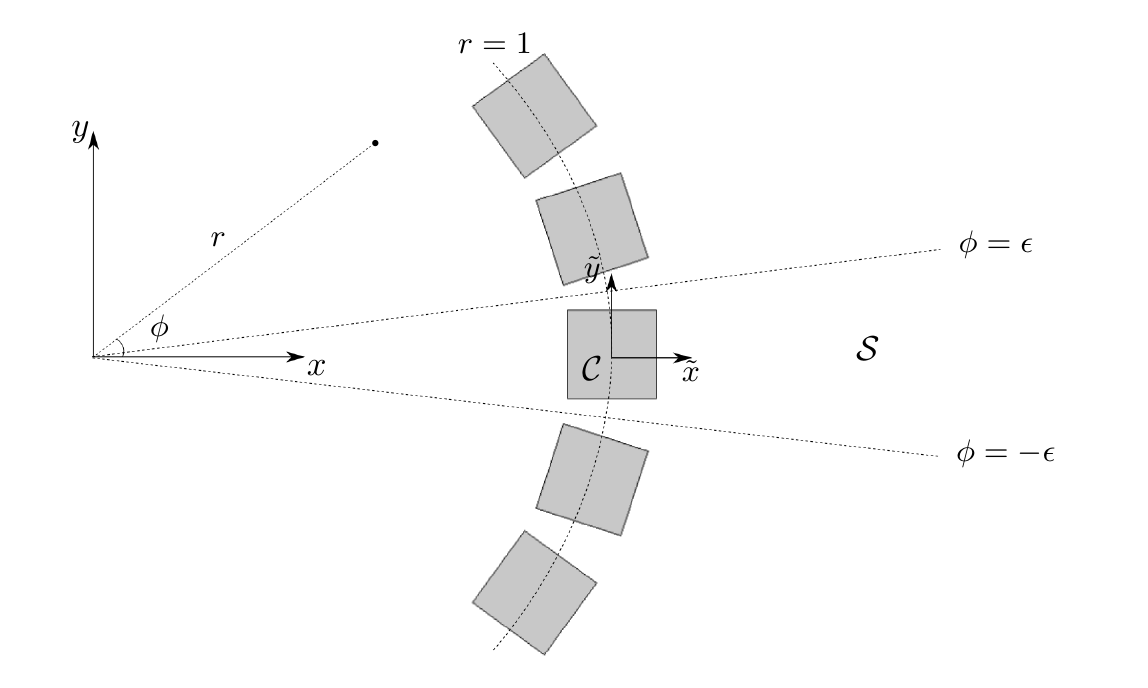}
\caption{Geometry of the problem in the case of square inclusions, showing the elementary cell $\mathcal{S}$.}
\label{fig:WB_geom}
\end{figure}
\noindent We consider an array of $N$ identical bounded inclusions located periodically around a circular ring of radius $r_0$; defining $\epsilon=\pi/N$ as half of the angular period of the array provides an intuitive small parameter. Dimensionless Cartesian co-ordinates $(x,y)$ are defined with respect to the centre of the ring such that the radius is scaled to 1, and the associated polar co-ordinates are $(r,\phi)$. A second set of Cartesian co-ordinates $(\tilde{x},\tilde{y})$, which are scaled and shifted, are defined via $x=1+\epsilon\tilde{x}$, $y=\epsilon\tilde{y}$, and using these we define the boundary $\partial\mathcal{C}$ of an arbitrarily-chosen inclusion $\mathcal{C}$ to satisfy the equation $f(\tilde{x},\tilde{y})=0$. An elementary cell $\mathcal{S}$ is chosen as the infinite wedge $|\phi|<\epsilon$, which we assume contains only the inclusion $\mathcal{C}$, as shown in figure \ref{fig:WB_geom}, and given the inherent periodicity, along with Bloch's theorem, solutions in this wedge-shaped cell can be used to generate the full field everywhere.

Assuming time dependence $\exp({-i\omega t})$, we seek radiating solutions of the planar Helmholtz eigenvalue problem in $\mathcal{S}$:
\begin{equation}\label{helm}
\left(\epsilon^2\nabla^2+\Omega^2\right)u(\mathbf{x})=0,
\end{equation}
where $\Omega=\epsilon\omega  r_0/{c}$ is the dimensionless frequency, $c$ the wave speed, and the Laplace operator is defined in terms of the dimensionless co-ordinates $(x,y)$. The homogeneous Neumann condition 
\begin{equation}\label{neumann}
\frac{\partial u}{\partial n }=0
\end{equation}
is imposed on the inclusion boundary $\partial\mathcal{C}$,
along with angular quasi-periodicity across the cell:
\begin{equation}\label{bloch}
u\big|_{\phi=\epsilon}=e^{2i\beta}u\big|_{\phi=-\epsilon}, \hspace{1cm}\frac{\partial u}{\partial\phi}\bigg|_{\phi=\epsilon}=e^{2i\beta}\frac{\partial u}{\partial\phi}\bigg|_{\phi=-\epsilon},
\end{equation}
where 
\begin{equation}\label{beta def}
\beta=\frac{\pi}{2}-\epsilon m
\end{equation}
is required by cyclic continuity for $m\in\mathbb{Z}$.

\section{Exponentially small curvature-induced damping}\label{sec:Qfactor}
Resonances of open systems, sometimes referred to as quasi-normal modes~\cite{ching98a}, are characterised by complex eigenfrequencies. A critical figure of merit, that emphasises how such solutions correspond to physical, time-dependent fields, is the Q-factor, given by 
\begin{equation}\label{q}
\text{Q-factor} \equiv
\frac{\text{maximum energy stored in cycle}}{\text{energy radiated per radian of cycle}}\sim\frac{\operatorname{Re}(\omega)}{|2\operatorname{Im}(\omega)|},
\end{equation}
where the two expressions coincide in the prevalent limit $\operatorname{Im}(\omega)/\operatorname{Re}(\omega)\to0$ \cite{Collin}.
 
In this paper we set out to analyse the limiting Q-factor that, given in terms of the dimensionless frequency by
\begin{equation}\label{Qlim}
\mathcal{Q}={\operatorname{Re}(\Omega)}/{|2\operatorname{Im}(\Omega)|},
\end{equation}
for resonances governed by the eigenvalue problem of section \ref{sec:formulation}, in the asymptotic limit $\epsilon\to0$, and as a function of the inclusion geometry and the mode number $m$. The asymptotic smallness of $\operatorname{Im}(\Omega)/\operatorname{Re}(\Omega)$ in this limit will provide a posteriori justification for definition \eqref{Qlim}. In fact, for moderate $m$ we shall actually find $\operatorname{Im}(\Omega)$ to be \textit{exponentially} small in $\epsilon$, with $\operatorname{Re}(\Omega)$ of course approaching the corresponding Rayleigh--Bloch frequency; the limiting Q-factor, $\mathcal{Q}$, is accordingly exponentially large. To help guide the asymptotic analysis of the next section, it is useful to intuitively describe the physical mechanism for curvature-induced radiation loss. Whereas the essential physics are akin to curved wave guides and micro-ring resonators, here the short-scale cyclic periodicity and particularly the $\epsilon$-quantisation in the Bloch conditions \eqref{bloch} modifies the traditional physical picture.

\begin{figure}[t]
  \centering
  \setlength{\unitlength}{\textwidth}
\includegraphics[width=.75\linewidth]{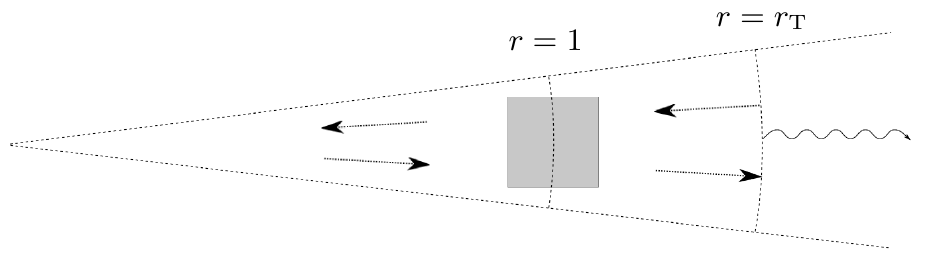}
\caption{Illustration of the radial field dependence on the interior and exterior of the ring, where the dotted arrows represent evanescent fields decaying in the directions indicated. As $r\to1$, the components of the field decaying toward the inclusion are exponentially smaller than those decaying away from the inclusion.}
\label{fig:waves}
\end{figure}

For a Rayleigh--Bloch wave guided by a linear array of inclusions, the wave field attenuation transverse to the array is exponential. Since for $\epsilon\ll1$ the ring is only slightly curved relative to the linear array, it is plausible to think this is also the case for a structured ring. However, for a finite system we expect outward radiation of energy, and hence at some larger radius the field must propagate energy and accordingly attenuate algebraically in the radial direction (see Fig.~\ref{fig:waves}). The exponential smallness of the outward radiation is essentially determined by the radial `tunnelling' distance from the structured ring over which the wave field remains evanescent.  We can estimate this distance by envisaging the unit cell $\mathcal{S}$ as a `virtual' waveguide whose $O(\epsilon)$ thickness slowly grows linearly with $r$, with boundary conditions that are anti-periodic to leading order as deduced from \eqref{bloch} and \eqref{beta def}. The evanescent field attenuating away from the inclusion excites the most slowly-decaying modes of this waveguide, and considering the waveguide to be locally straight, one readily finds an approximate cut-off radius $r_{\text{T}}={\pi}/({2\Omega_0})$, where $\Omega_0$ denotes the Rayleigh--Bloch frequency.

\begin{figure}[t]
  \centering
  \setlength{\unitlength}{\textwidth}
\includegraphics[width=.8\linewidth]{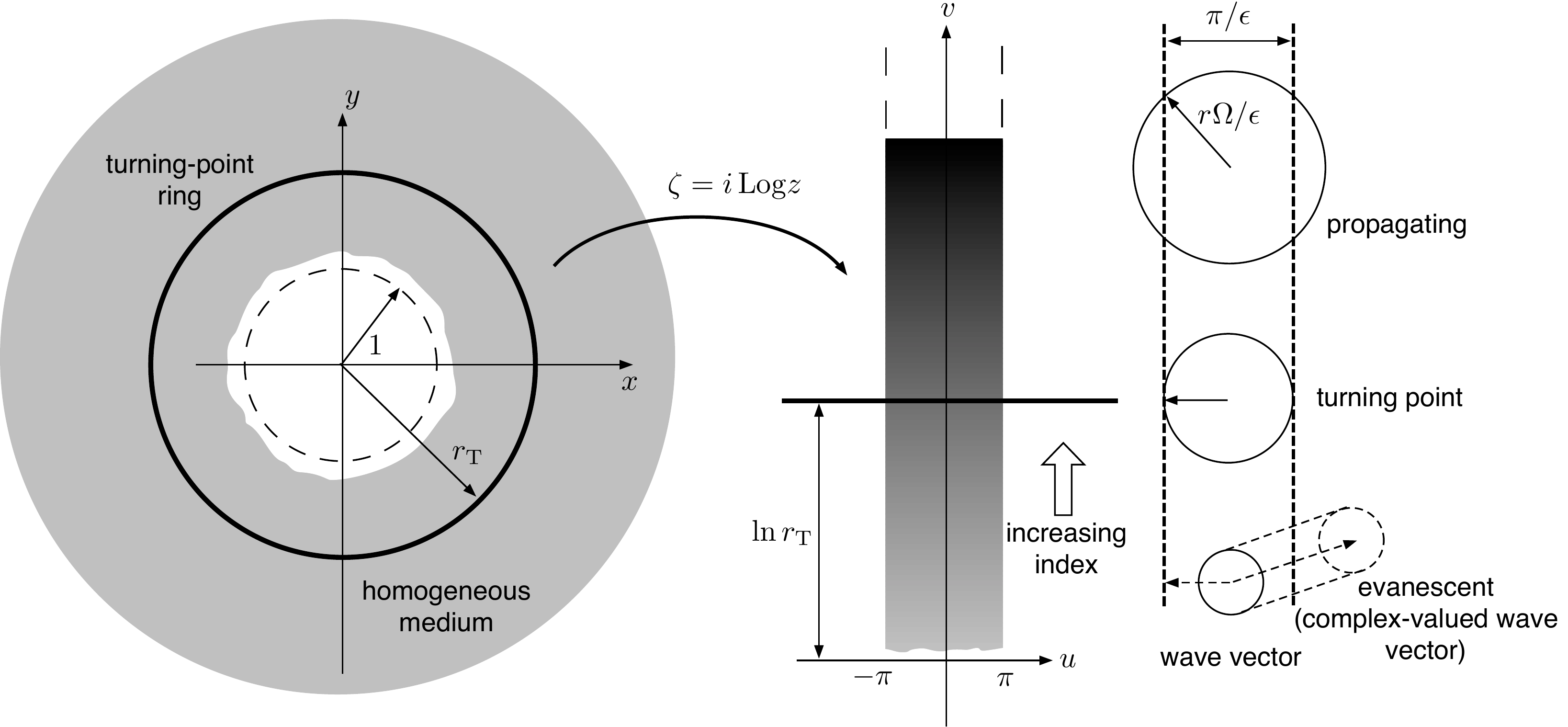}
\caption{Conformal mapping of the domain external to the curved ring to a Cartesian semi-infinite strip. In the transformed plane, the Rayleigh--Bloch wave guided along the horizontal axis transversely attenuates evanescently into a material whose effective index grows with $v=\ln r$; at $r\approx r_{\text{T}}=\pi/(2\Omega)$ the wave field begins to propagate.}
\label{fig:conformal}
\end{figure}

Curvature-induced loss can alternatively be understood with the help of a conformal mapping. A classical technique for analysing curved waveguides is to map the waveguide and its surrounding into an auxiliary plane where the waveguide is straight, with the consequence of distorting the spatial distribution of the material index \cite{Heiblum:75}. While it would be technically difficult to directly apply this technique for the inclusion-ring geometry, for the present qualitative discussion it is sufficient to consider the mapping of the region $r-1\gg O(\epsilon)$ external to the ring, where the wave field is governed by the constant-index Helmholtz equation \eqref{helm} (see Fig.~\ref{fig:conformal}).  Defining the complex variable $z=x+iy=r\exp(i\phi)$, and the auxiliary complex variable $\zeta=u+iv$, the mapping $\zeta=i\text{Log(z)}=i\ln r-\phi$, where $-\pi< \phi< \pi$, takes the physical domain $r>1$ to the semi-infinite strip ${|u|<\pi,v>0}$. Writing $u(x,y)=w(u,v)$, it is readily verified that the Helmholtz equation \eqref{helm} transforms to 
\begin{equation}\label{helm zeta}
\pd{^2w}{u^2}+\pd{^2w}{v^2}+[r(v)\Omega/\epsilon]^2 w=0,
\end{equation}
where $r(v)=\exp(v)$. Note that instead of choosing a specific branch of the logarithm we can consider a mapping to the entire half-plane $v>0$, with the transformed wave field $2\pi$ periodic as a function of $u$; the latter periodicity condition is a manifestation of the cyclic quantisation of the allowed linear-array Rayleigh--Bloch frequency. Eq.~\eqref{helm zeta} shows that in the transformed plane the effective index is no longer homogeneous, but grows with $v$; thus the  evanescent wave associated with the Rayleigh--Bloch wave attenuates into a material whose index increases until eventually the wave field becomes leaky (positioning a high-index material in the vicinity of an interface supporting surface waves is in fact a well-known experimental technique for phase-matching bulk and surface waves, see Ref.~\cite{Maier:07}). Specifically, for $r=O(1)$ a leading-order ray-optics approximation implies $|\mathbf{k}|^2\approx (r\Omega)^2/\epsilon^2$, where $\mathbf{k}$ denotes a normalised wave vector in the transformed plane.  But the Bloch conditions \eqref{bloch} suggest a leading-order lower bound, $\epsilon^{-1}\pi/2$, on the projection of $\mathbf{k}$ in the direction of the $u$ axis. Accordingly, the propagation requirement of a real wave vector yields, once again, $r>\pi/(2\Omega_0)$.

More accurately, noting that \eqref{helm zeta} is separable, and given the Bloch conditions \eqref{bloch}, we can write $w=\exp[i(\mp\pi/2+\epsilon m)u/\epsilon]W^{\pm}[v(r)]$ plus orthogonal terms of higher azimuthal order. Substitution into \eqref{helm zeta} shows that 
\begin{equation}\label{r dep}
\epsilon^2\frac{d^2W^{\pm}}{dv^2}+\left[r^2\Omega^2-(\pm\pi/2-\epsilon m)^2\right] W^{\pm}=0.
\end{equation}
Assuming the expansion $\Omega\sim \Omega_0 + \epsilon \Omega_1+ O(\epsilon^2)$, this lowest-order mode is radially propagating if
\begin{equation}\label{estimate}
r > \frac{\pi}{2\Omega_0}\mp \epsilon\frac{ m }{\Omega_0}-\epsilon\frac{\pi\Omega_1}{2\Omega_0^2} + O(\epsilon^2);
\end{equation}
higher-order azimuthal components `turn-on' at larger distances. Of course the same conclusion can be deduced by an analogous separation of variables in polar coordinates in the context of the above-discussed wedge-shaped waveguide. The more accurate estimate \eqref{estimate} suggests that the turning point bifurcates, slightly, with respect to the wave-field components propagating clockwise and anticlockwise, the relative strength of these contributions being dictated by the evanescent tail of the quasi-guided Rayleigh--Bloch waves along the ring, and hence by the geometry of the inclusions. Notwithstanding the smallness of this bifurcation we shall find it important when calculating the wave field and the exponentially small damping to leading order. 

\section{Asymptotic analysis} \label{sec:analysis}
\subsection{Inner region}\label{sssec:inner}
\begin{figure}[ht]
  \centering
  \setlength{\unitlength}{\textwidth}
\includegraphics[width=\linewidth]{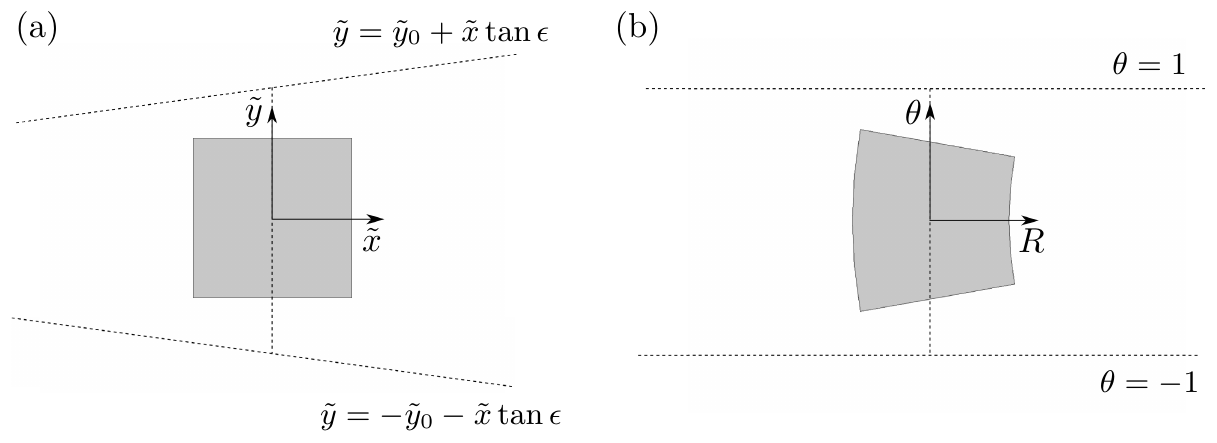}
\caption{Schematic of the inner region viewed in (a) shifted, scaled Cartesian co-ordinates and (b) shifted, scaled polar co-ordinates illustrated for the case of a square inclusion. The inclusion boundary $\partial\mathcal{C}$ is defined in Cartesian co-ordinates by $f(\tilde{x},\tilde{y})=0$, and we have defined $\tilde{y}_0=(1/\epsilon)\tan\epsilon$.}
\label{fig:inner}
\end{figure}

\noindent Assuming that the inclusions are contained within an $O(\epsilon)$ annulus  around the ring $r=1$, we define shifted, scaled polar co-ordinates $(R,\theta)$ inside the cell via the relations $r=1+\epsilon R$ and $\phi=\epsilon\theta$, giving rise to an inner region in which $R,\theta=O(1)$. In this region we seek an inner expansion $u(\mathbf{x})=\Phi(R,\theta)$ that satisfies
\begin{equation}\label{helm_asy}
\left(\frac{\partial^2}{\partial R^2}+\frac{\epsilon}{1+\epsilon R}\frac{\partial}{\partial R}+\frac{1}{(1+\epsilon R)^2}\frac{\partial^2}{\partial\theta^2}+\Omega^2\right)\Phi(R,\theta)=0,
\end{equation}
with quasi-periodicity conditions from (\ref{bloch}), expanded as
\begin{equation}\label{quasi_phi}\begin{split}
\Phi(R,1)&=(-1+2i\epsilon m+\dots)\Phi(R,-1),\\
\frac{\partial\Phi}{\partial\theta}(R,1)&=(-1+2i\epsilon m+\dots)\frac{\partial\Phi}{\partial\theta}(R,-1),
\end{split}\end{equation}
and $\Phi(R,\theta)$ decaying as $R\to\pm\infty$. As for the Neumann condition \eqref{neumann}, we note that the mapping from Cartesian to polar co-ordinates results in an asymptotically small distortion of the inclusion $\mathcal{C}$, as illustrated in figure \ref{fig:inner}(b). Consequently, applying \eqref{neumann} generally entails deriving asymptotically equivalent conditions on the `undeformed' boundary $f(R,\theta)=0$.

\begin{figure}[t]
  \centering
  \setlength{\unitlength}{\textwidth}
\includegraphics[width=.6\linewidth]{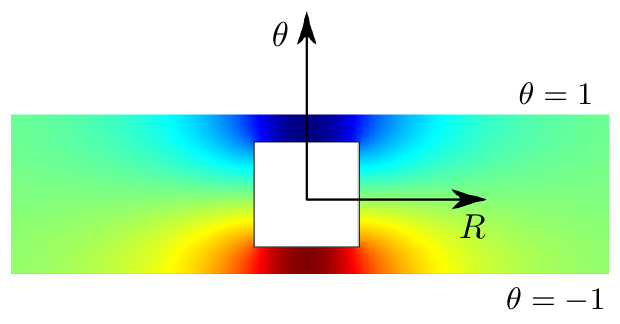}
\caption{Leading order inner expansion $\Phi_0(R,\theta)$ for a square inclusion of side $1.3\epsilon$, which in these co-ordinates is identical to a standing Rayleigh--Bloch wave for the linear array.}
\label{fig:inner_o1}
\end{figure}

The forms of (\ref{helm_asy}) and (\ref{quasi_phi}) lead us to pose an ansatz of the form
\begin{equation}\label{ansatz}
\Phi(R,\theta)\sim\Phi_0(R,\theta)+\epsilon\Phi_1(R,\theta)+\dots,\hspace{.5cm}
\Omega\sim\Omega_0+\epsilon\Omega_1+\dots,
\end{equation}
which yields an eigenvalue problem for the leading order terms:
\begin{equation}\label{O1}
\left(\frac{\partial^2}{\partial R^2}+\frac{\partial^2}{\partial \theta^2}+\Omega_0^2\right)\Phi_0(R,\theta)=0,
\end{equation}
subject to the anti-periodic boundary conditions
\begin{equation}\label{anti_phi}
\Phi_0(R,1)=-\Phi_0(R,-1), \hspace{1cm}\frac{\partial\Phi_{0}}{\partial\theta}(R,1)=-\frac{\partial\Phi_{0}}{\partial\theta}(R,-1),
\end{equation}
and the homogeneous Neumann condition
\begin{equation}\label{neumann_phi}
\frac{\partial\Phi_0}{\partial N}=0
\end{equation}
on the nominal boundary $f(R,\theta)=0$, where the derivative is in the direction of the normal $\mathbf{N}$ to this boundary. The decay condition then identifies the solution as a standing Rayleigh--Bloch wave for the linear array (see figure \ref{fig:inner_o1}), and $\Omega_0\in\mathbb{R}$ is the corresponding eigenfrequency.

In the limit $R\to\infty$, the solution $\Phi_0(R,\theta)$ is comprised of a linear combination of evanescent waveguide-type modes. These are separated in magnitude by exponential order and hence to leading exponential order only the most slowly-decaying one needs to be considered for matching. Assuming $\Phi_0(R,\theta)$ is chosen to be real, we have
\begin{equation}\label{outer_inner}
\Phi_0(R,\theta)\sim A\exp\left(-R\sqrt{\left(\frac{\pi}{2}\right)^2-\Omega_0^2}\right)\sin\left(\frac{\pi}{2}\theta+\alpha\right)
\end{equation}
as $R\to\infty$, where the constants $\alpha,A\in\mathbb{R}$, along with the eigenfrequency $\Omega_0$, are straightforward to extract from a numerical solution; here we utilise standard finite element solvers to extract these.

In general it is necessary to proceed to the next order in the asymptotic hierarchy to calculate the frequency correction $\Omega_1$, as this term can be shown to affect the leading-order solution in the outer region. 
If the undeformed inclusions are fore-aft symmetric, however, $\Omega_1$ vanishes as we shall now show, and hence we do not require any further analysis of the inner problem. To see this, consider the inner region in Cartesian co-ordinates as shown in figure \ref{fig:inner}(a), assuming that the inclusion is symmetric about the $\tilde{y}$-axis. 
Writing $u(\mathbf{x})=w(\tilde{x},\tilde{y})$ in this region, we have 
\begin{equation}\label{tilde_equation}
\left(\frac{\partial^2}{\partial\tilde{x}^2}+\frac{\partial^2}{\partial\tilde{y}^2}+\Omega^2\right)w(\tilde{x},\tilde{y})=0,
\end{equation}
subject to the homogeneous Neumann condition on $f(\tilde{x},\tilde{y})=0$, attenuation as $\tilde{x}\to\pm\infty$, and quasi-periodicity across the domain. Due to the symmetry of the inclusion, taking $\epsilon\to-\epsilon$ is equivalent to a reflection of the geometry in the $\tilde{y}$-axis, along with complex conjugation of the Bloch factor appearing in the boundary conditions. Under this transformation, the inner solution is thus given by $\bar{w}(-\tilde{x},\tilde{y})$, and the corresponding frequency is $\bar{\Omega}$, where the bar denotes quantities associated with the corresponding adjoint problem. As we shall see, the imaginary part of the frequency is exponentially small with respect to $\epsilon$, so we deduce that the substitution $\epsilon\to-\epsilon$ has no effect on the eigenfrequency $\Omega$ to all algebraic orders in $\epsilon$. With this in mind, posing the expansion $\Omega\sim\Omega_0+\epsilon\Omega_1+\dots$ leads us to conclude that $\Omega_1=0$. From hereon we will restrict ourselves to fore-aft symmetric inclusions so that this is guaranteed to be the case. We note that for an inclusion of arbitrary shape the frequency correction $\Omega_1$ can be calculated by deriving a solvability condition on the $O(\epsilon)$ inner-region problem, which involves an altered Neumann boundary condition as provided by appendix A of \cite{Hewett_16}. 

\subsection{Outer region}\label{sssec:outer} 

We now turn our attention to the outer region, which lies exterior to the ring and in which $r-1=O(1)$.  In this region, we define the field $u(\mathbf{x})=U(r,\theta)$, which satisfies the equation

\begin{equation}\label{helm_scaled}
\left(\epsilon^2\frac{\partial^2}{\partial r^2}+\epsilon^2\frac{1}{r}\frac{\partial}{\partial r}+\frac{1}{r^2}\frac{\partial^2}{\partial\theta^2}+\Omega^2\right)U(r,\theta)=0,
\end{equation}
subject to quasi-periodicity,
\begin{equation}\label{quasi_U}\begin{split}
U(r,1)&=(-1+2i\epsilon m+\dots)U(r,-1),\\
\frac{\partial U}{\partial\theta}(r,1)&=(-1+2i\epsilon m+\dots)\frac{\partial U}{\partial\theta}(r,-1),
\end{split}\end{equation}
and matching with $\Phi(R,\theta)$ as $r\to1$. We seek a general solution using the WKB ansatz:
\begin{equation}\label{WKB}
U(r,\theta)\sim e^{i\varphi(r)/\epsilon}\left(U_{0}(r,\theta)+\epsilon U_{1}(r,\theta)+\dots\right).
\end{equation} 
The resulting leading order problem is given by 
\begin{equation}\label{WKB_1}
\left(\frac{\partial^2}{\partial\theta^2}+\lambda^2\right)U_{0}(r,\theta)=0,
\end{equation}
where 
\begin{equation}\label{lambda}
\lambda^2=r^2\left[\Omega_0^2-\left(\frac{\mathrm{d}\varphi}{\mathrm{d}r}\right)^2\right],
\end{equation}
subject to the anti-periodic boundary conditions
\begin{equation}\label{anti}
U_{0}(r,1)=-U_{0}(r,-1), \hspace{1cm}\frac{\partial U_{0}}{\partial\theta}(r,1)=-\frac{\partial U_{0}}{\partial\theta}(r,-1).
\end{equation}
This problem has an infinite number of independent solutions corresponding to $\lambda=\pm(2n+1/2)\pi$, $n\in\mathbb{Z}$, but matching with (\ref{outer_inner}) implies that only those with $\lambda=\pm\pi/2$ appear to leading exponential order. The corresponding solutions are given by 
\begin{equation}\label{wkb_u}
U_0(r,\theta)=U_{0+}(r,\theta)+U_{0-}(r,\theta),
\end{equation}
where 
\begin{equation}\label{WKB_sol}
U_{0\pm}(r,\theta)=F_{\pm}(r)e^{\pm i\pi\theta/2}.
\end{equation}
At the next order, we have the equation
\begin{equation}\label{WKB_e}
\left(\frac{\partial^2}{\partial\theta^2}+\lambda^2\right)U_{1}=-\left(2ir^2\frac{\mathrm{d}\varphi}{\mathrm{d}r}\frac{\partial}{\partial r}+ir^2\frac{\mathrm{d}^2\varphi}{\mathrm{d}r^2}+ir\frac{\mathrm{d}\varphi}{\mathrm{d}r}\right)U_{0},
\end{equation}
where the boundary conditions at this order,
\begin{equation}\label{anti_m}\begin{split}
U_{1}(r,1)&=-U_{1}(r,-1)+2imU_{0}(r,-1),\\
\frac{\partial U_{1}}{\partial\theta}(r,1)&=-\frac{\partial U_{1}}{\partial\theta}(r,-1)+2im\frac{\partial U_{0}}{\partial\theta}(r,-1),
\end{split}\end{equation}
depend on the mode number $m$.
For each of the solutions in (\ref{wkb_u}) we derive a solvability condition using the Fredholm alternative: we subtract the product of (\ref{WKB_e}) with $U_{0\pm}(r,\theta)$ from the product of $\eqref{WKB_1}$, written for $U_{0\pm}(r,\theta)$, with $U_{1}(r,\theta)$ and integrate over the angular variable $\theta$. After applying the boundary conditions, we are left with equations for $F_\pm(r)$, given by
\begin{equation}\label{AB_eqn}
r\frac{\mathrm{d}}{\mathrm{d}r}\left(F_{\pm}^2r\frac{\mathrm{d}\varphi}{\mathrm{d}r}\right)=\pm im\pi F_{\pm}^2,
\end{equation}
which are straightforward to solve. 

In agreement with the discussion of section \ref{sec:Qfactor}, we deduce from (\ref{lambda}) that the phase function $\varphi(r)$ changes from real to imaginary at a turning-point radius $r=r_\text{T}$ where
\begin{equation}\label{rt}
r_\text{T}\equiv\frac{\pi}{2\Omega_0}, 
\end{equation}
which we know is greater than 1. It follows from \eqref{rt} that if the Rayleigh--Bloch frequency $\Omega_0$ is well below the band-edge light-line frequency $\pi/2$, the turning point lies in the outer region, i.e. $r_\text{T}-1=O(1)$. Conversely, our asymptotic analysis breaks down as $r_\text{T}\to1$, in a manner briefly discussed in section \ref{sec:delta}.

The WKB solution now follows from \eqref{wkb_u} together with the appropriate solutions of \eqref{AB_eqn}. In the region $1<r<r_\text{T}$ the field is evanescent, and is comprised of two outward-decaying terms and two inward-decaying terms:
\begin{equation}\label{ue}\begin{split}
U(r,\theta)\sim\left(r_\text{T}^2-r^2\right)^{-1/4}&\Big[\left\{U_{0+}(r,\theta)+U_{0-}(r,\theta)\right\}e^{-\psi(r)/\epsilon}\\
&+\left\{V_{0+}(r,\theta)+V_{0-}(r,\theta)\right\}e^{\psi(r)/\epsilon}\Big],
\end{split}\end{equation}
where 
\begin{align}\label{U_sol}
U_{0\pm}(r,\theta)&=B_{\pm}e^{\mp m h(r)}e^{\pm i\pi\theta/2},\\
V_{0\pm}(r,\theta)&=C_{\pm}e^{\pm m h(r)}e^{\pm i\pi\theta/2},\label{V_sol}
\end{align}
with
\begin{equation}\label{psi}
\psi(r)=\Omega_0\int_1^r(r_\text{T}^2/v^2-1)^{1/2}\mathrm{d}v,
\end{equation}
and
\begin{equation}\label{h}
h(r)=r_\text{T}\int_1^rv^{-2}(r_\text{T}^2/v^2-1)^{-1/2}\mathrm{d}v.
\end{equation}
In the following section we will find that in a small region near the turning point the inward-decaying terms in (\ref{ue}) are comparable in magnitude to the outward-decaying terms. This means that in the majority of the evanescent region, including the region in which the inner and outer solutions must match, the former are exponentially small and can thus be neglected. We then find that the inner limit of the outward-decaying field matches with the outer limit (\ref{outer_inner}) of the leading-order inner expansion, provided that
\begin{equation}\label{match}
B_{\pm}=\pm\frac{1}{2i}A(r_\text{T}^2-1)^{1/4}e^{\pm i\alpha}.
\end{equation}
 In the region $r>r_\text{T}$ we have two outward-propagating terms,
\begin{equation}\label{ue2}
U(r,\theta)\sim\left(r^2-r_\text{T}^2\right)^{-1/4}\left\{W_{0+}(r,\theta)+W_{0-}(r,\theta)\right\}e^{i\varphi(r)/\epsilon},
\end{equation}
where
\begin{equation}\label{U_sol_2}W_{0\pm}(r,\theta)=D_{\pm}e^{\pm im p(r)}e^{\pm i\pi\theta/2},
\end{equation}
with
\begin{equation}\label{phi}
\varphi(r)=\Omega_0\int_{r_\text{T}}^r(1-r_\text{T}^2/v^2)^{1/2}\mathrm{d}v,
\end{equation}
and
\begin{equation}
p(r)=\operatorname{arctan}\left\{(r^2/r_\text{T}^2-1)^{1/2}\right\}.
\end{equation}
In the following section we will establish connection formulae that lead to the following expression for the amplitudes of the exponentially small outgoing waves:
\begin{equation}\label{D}
D_\pm=\pm2^{-13/12}A(r_\text{T}^2-1)^{1/4}e^{-\left\{\psi(r_\text{T})\pm\epsilon mh(r_\text{T})\right\}/\epsilon}e^{i(\pm\alpha-\pi/4)},
\end{equation}
where the constants $\alpha$, $A$ were introduced in (\ref{outer_inner}).

\subsection{Transition region}\label{ssssec:connection}
The leading order WKB approximation in the outer region suggests a turning point at $r=r_\text{T}$. The standard turning-point analysis of WKB theory leads us to introduce a scaled co-ordinate variable $s=(r-r_{\text{T}})/\epsilon^{2/3}$, 
and then seek an expansion for $U(r,\theta)$ whose terms are increments in powers of $\epsilon^{2/3}$. Care must be taken, however, as for non-zero $m$ the boundary conditions (\ref{quasi_U}) can only be satisfied if there is also a term at order $\epsilon$ beyond the leading order, which in turn implies that there must be a term at order $\epsilon^{1/3}$, and hence our ansatz needs to be adjusted. The reason for the failure has been set out in section \ref{sec:Qfactor}: 
For each of the outward-decaying terms in (\ref{ue}), the `actual' turning point is shifted from $r_\text{T}$ by an $O(\epsilon)$ distance, and hence a naive expansion about $r_\text{T}$ leads to a solution that is not uniformly asymptotic. Based on this understanding, a natural way to proceed is to utilise the method of strained co-ordinates: appealing to the linearity of the problem we seek a transition region solution at leading exponential order of the form 
\begin{equation}
U(r,\theta)=\epsilon^{-1/6}\left\{G_+(s_+,\theta)+G_-(s_-,\theta)\right\},
\end{equation}
where
\begin{equation}\label{transition_A}
G_{\pm}(s_\pm,\theta)\sim G_{0\pm}(s_\pm,\theta)+\epsilon^{2/3}G_{1\pm}(s_\pm,\theta)+\epsilon G_{2\pm}(s_\pm,\theta)+\dots,
\end{equation}
with two different strained co-ordinates $s_{\pm}$ defined via the expansions $r\sim r_\text{T}+\epsilon^{2/3}s_{\pm}+\epsilon r_{1\pm}+\dots$.
The corrections $r_{1\pm}$ are to be chosen such that the expansions are uniformly asymptotic as $\epsilon\to0$. Substituting into equation (\ref{helm_scaled}) leads to the same problem for $G_{0+}$ and $G_{0-}$, consisting of
\begin{equation}\label{O1_F}
\left(\frac{\partial^2}{\partial\theta^2}+r_\text{T}^2\Omega_0^2\right)G_{0\pm}(s_\pm,\theta)=0,
\end{equation}
subject to the anti-periodic boundary conditions
\begin{equation}\label{anti_H0}
G_{0\pm}(s_\pm,1)=-G_{0\pm}(s_\pm,-1), \hspace{.7cm}\frac{\partial G_{0\pm}}{\partial\theta}(s_\pm,1)=-\frac{\partial G_{0\pm}}{\partial\theta}(s_\pm,-1).
\end{equation}
Its solutions are given by
\begin{equation}\label{F_0}
G_{0\pm}(s_\pm,\theta)=H_{0\pm}(s_\pm)e^{\pm i\pi\theta/2}.
\end{equation}
At the next order, we have
\begin{equation}\label{Oe23_F}
\left(\frac{\partial^2}{\partial\theta^2}+r_\text{T}^2\Omega_0^2\right)G_{1\pm}(s_{\pm},\theta)=\left(\frac{2s_\pm}{r_\text{T}}\frac{\partial^2}{\partial\theta^2}-r_\text{T}^2\frac{\partial^2}{\partial s^2}\right)G_0(s_\pm,\theta)
\end{equation}
also subject to anti-periodic boundary conditions
\begin{equation}\label{anti_H1}
G_{1\pm}(s_\pm,1)=-G_{1\pm}(s_\pm,-1), \hspace{.7cm}\frac{\partial G_{1\pm}}{\partial\theta}(s_\pm,1)=-\frac{\partial G_{1\pm}}{\partial\theta}(s_\pm,-1).
\end{equation}
Solvability conditions at this order, derived by a method analogous to that used to derive (\ref{AB_eqn}), lead to Airy equations
\begin{equation}\label{airy}
\frac{\mathrm{d}^2H_{0\pm}}{\mathrm{d}s_\pm^2}+\frac{2\Omega_0^2}{r_\text{T}}s_\pm H_{0\pm}=0,
\end{equation}
which have the solutions 
\begin{equation}\label{airy_A}
H_{0\pm}(s_\pm)=a_\pm\operatorname{Ai}\left(-\left(2\Omega_0^2/r_\text{T}\right)^{1/3}s_\pm\right)+b_\pm\operatorname{Bi}\left(-\left(2\Omega_0^2/r_\text{T}\right)^{1/3}s_\pm\right),\end{equation}
where $\operatorname{Ai}$, $\operatorname{Bi}$ are Airy functions of the first and second kinds respectively. The explicit form of the strained co-ordinates $s_\pm$ are still to be determined. In order to do so, we proceed to the next order problem, consisting of
\begin{equation}\label{Oe_F}
\left(\frac{\partial^2}{\partial\theta^2}+r_\text{T}^2\Omega_0^2\right)G_{2\pm}(s_\pm,\theta)=\left(\frac{2r_{1\pm}}{r_\text{T}}\frac{\partial^2}{\partial\theta^2}-2r_\text{T}^2\Omega_0\Omega_1\right)G_{0\pm}(s_\pm,\theta),
\end{equation}
subject to the $m$-dependent boundary conditions
\begin{equation}\label{anti_m_G2}\begin{split}
G_{2\pm}(s_\pm,1)&=-G_{2\pm}(s_\pm,-1)+2imG_{0\pm}(s_\pm,-1),\\
\frac{\partial G_{2\pm}}{\partial\theta}(s_\pm,1)&=-\frac{\partial G_{2\pm}}{\partial\theta}(s_\pm,-1)+2im\frac{\partial G_{0\pm}}{\partial\theta}(s_\pm,-1).
\end{split}\end{equation}
The solvability conditions at this order yield
\begin{equation}\label{r0}
r_{1\pm}=\mp m/\Omega_0,
\end{equation}
which agrees with our preliminary estimate \eqref{estimate} for the bifurcation of the turning-point radius.

\subsubsection{Connection formulae}\label{sssssec:connect}
By considering intermediate regions of order $\epsilon^\nu$ either side of the turning point, where $0<\nu<2/3$, the WKB solutions (\ref{ue}) and (\ref{ue2}) are matched with the transition region solution (\ref{airy_A}) to leading order, yielding connection formulae between their respective preceding constants. On the evanescent side of the turning point, we find
\begin{align}\label{connection1}
a_\pm&=2^{13/12}\pi^{1/6}\Omega_0^{1/2}C_\pm e^{-\{\psi(r_\text{T})\pm \epsilon mh(r_\text{T})\}/\epsilon},\\
b_\pm&=2^{1/12}\pi^{1/6}\Omega_0^{1/2}B_\pm e^{\{\psi(r_\text{T})\pm \epsilon mh(r_\text{T})\}/\epsilon},\label{connection2}
\end{align}
whilst on the propagating side we find
\begin{align}\label{aib}
a_\pm&=ib_\pm,\\
D_\pm&=2^{-1/6}\pi^{-1/6}\Omega_0^{-1/2}e^{i\pi/4}b_{\pm}.\label{connection}
\end{align}
Combining these results with (\ref{match}) leads to the expression (\ref{D}) for the amplitudes of the exponentially small outgoing waves. Note also that (\ref{connection1}), (\ref{connection2}) and (\ref{aib}) together imply that the inward-decaying terms in (\ref{ue}) are exponentially smaller than the outward-decaying terms everywhere in the evanescent region where $r_\text{T}-r\gg\epsilon$. 

\begin{figure}[t]
  \centering
  \setlength{\unitlength}{\textwidth}
\includegraphics[width=\linewidth]{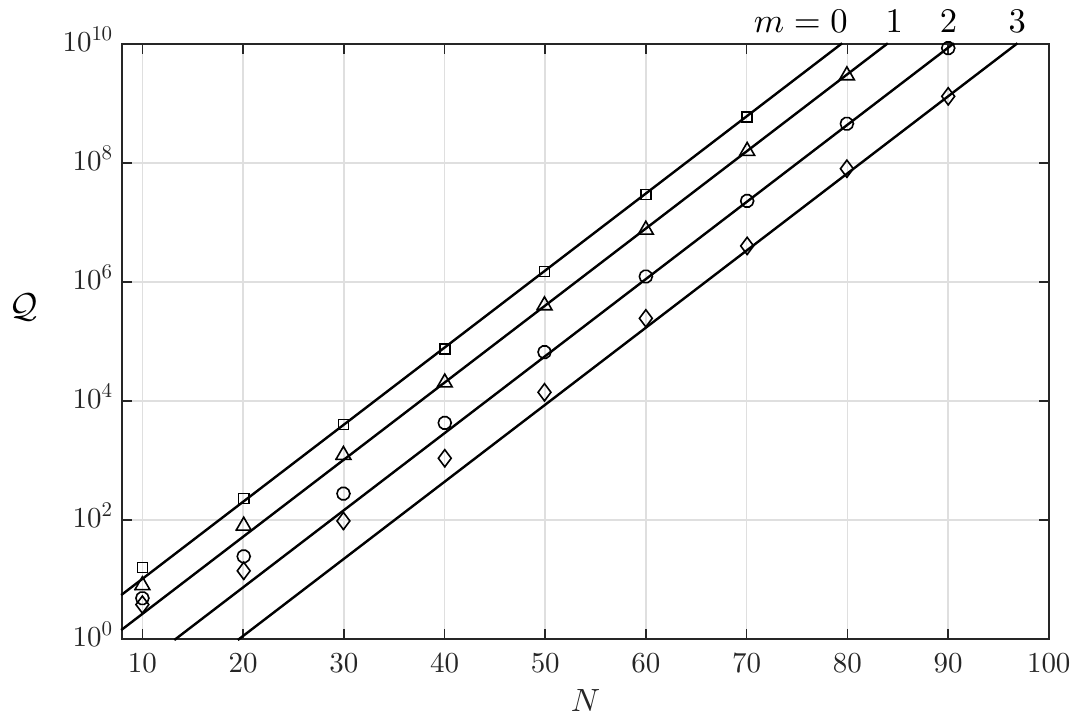}
\caption{Dependence of $\mathcal{Q}$ on $N$ for a ring of slit-like inclusions of length $1.2\epsilon$, corresponding to those in figure \ref{fig:RB}(b). Symbols are from full finite element simulation, with different symbols used for different values of $m$, and solid lines are from the asymptotic formula (\ref{Omi_final}). Note that the vertical axis is scaled logarithmically.}
\label{fig:Omi}
\end{figure}

\section{Radiation loss}\label{sec:damping}
In order to calculate the limiting Q-factor of the resonant system, we derive an equation that represents energy balance in the system; we subtract the product of (\ref{helm}) with the complex conjugate of ${u}(\mathbf{x})$ from the from the product of $u(\mathbf{x})$ with the complex conjugate of (\ref{helm}), and then integrate over a truncated wedge $\tilde{\mathcal{S}}$, which extends to a finite arc $r=\tilde{r}$ in the propagating region. The resulting equation is given by
\begin{equation}\label{energy}
\epsilon^2\int_{\partial\tilde{\mathcal S}}\left\{\bar{u}\frac{\partial u}{\partial r}-u\frac{\partial \bar{u}}{\partial r}\right\}\mathrm{d}l=-4i\Omega_\text{r}\Omega_\text{i}\int_{\tilde{\mathcal{S}}}\left|u\right|^2\mathrm{d}S.
\end{equation}
where $\Omega=\Omega_\text{r}+i\Omega_\text{i}$. Substituting the leading order outer and inner solutions into the left and right hand sides respectively, then expanding to leading order in $\epsilon$, we arrive at the following simple asymptotic expression:
\begin{equation}\label{Omi}
\Omega_\text{i}\sim\frac{-\left(|D_+|^2+|D_-|^2\right)}{\int\limits_{-1}^1\int\limits_{\infty}^{\infty}\left|\Phi_0(R,\theta)\right|^2\mathrm{d}R\mathrm{d}\theta}.
\end{equation}
Substituting (\ref{D}) for the constants $D_\pm$, and using the definition (\ref{Qlim}), we finally arrive at the result
\begin{equation}\label{Omi_final}
\mathcal{Q}\sim\frac{\Omega_0\int\limits_{-1}^1\int\limits_{\infty}^{\infty}\left|\Phi_0(R,\theta)\right|^2\mathrm{d}R\mathrm{d}\theta}{2^{5/12}(r_\text{T}^2-1)^{1/2}A^2}\operatorname{sech}\{2mh(r_\text{T})\}
e^{2\psi(r_\text{T})/\epsilon},
\end{equation}
where, using \eqref{psi} and \eqref{h}, respectively, 
\begin{equation}\label{psi and h}
\psi(r_{\text{T}})/\Omega_0=r_{\text{T}}\cosh^{-1} r_{\text{T}}-\sqrt{r_{\text{T}}^2-1}, \quad h(r_{\text{T}})=\ln\left(r_{\text{T}}+\sqrt{r_{\text{T}}-1}\right);
\end{equation}
recall also that $r_{\text{T}}=\pi/(2\Omega_0)$, and that $A$ is an $O(1)$ constant extracted from the numerical solution for the inner-region wave field $\Phi_0$ [cf.~\eqref{outer_inner}]; since $\Phi_0$ scales with $A$, $\mathcal{Q}$ is independent of the arbitrary magnitude of the resonance. Fig.~\eqref{fig:Omi} shows excellent agreement between \eqref{Omi_final} and finite-element simulations for a ring of slit-like inclusions.
\begin{figure}[t]
  \centering
  \setlength{\unitlength}{\textwidth}
\includegraphics[width=0.55\linewidth]{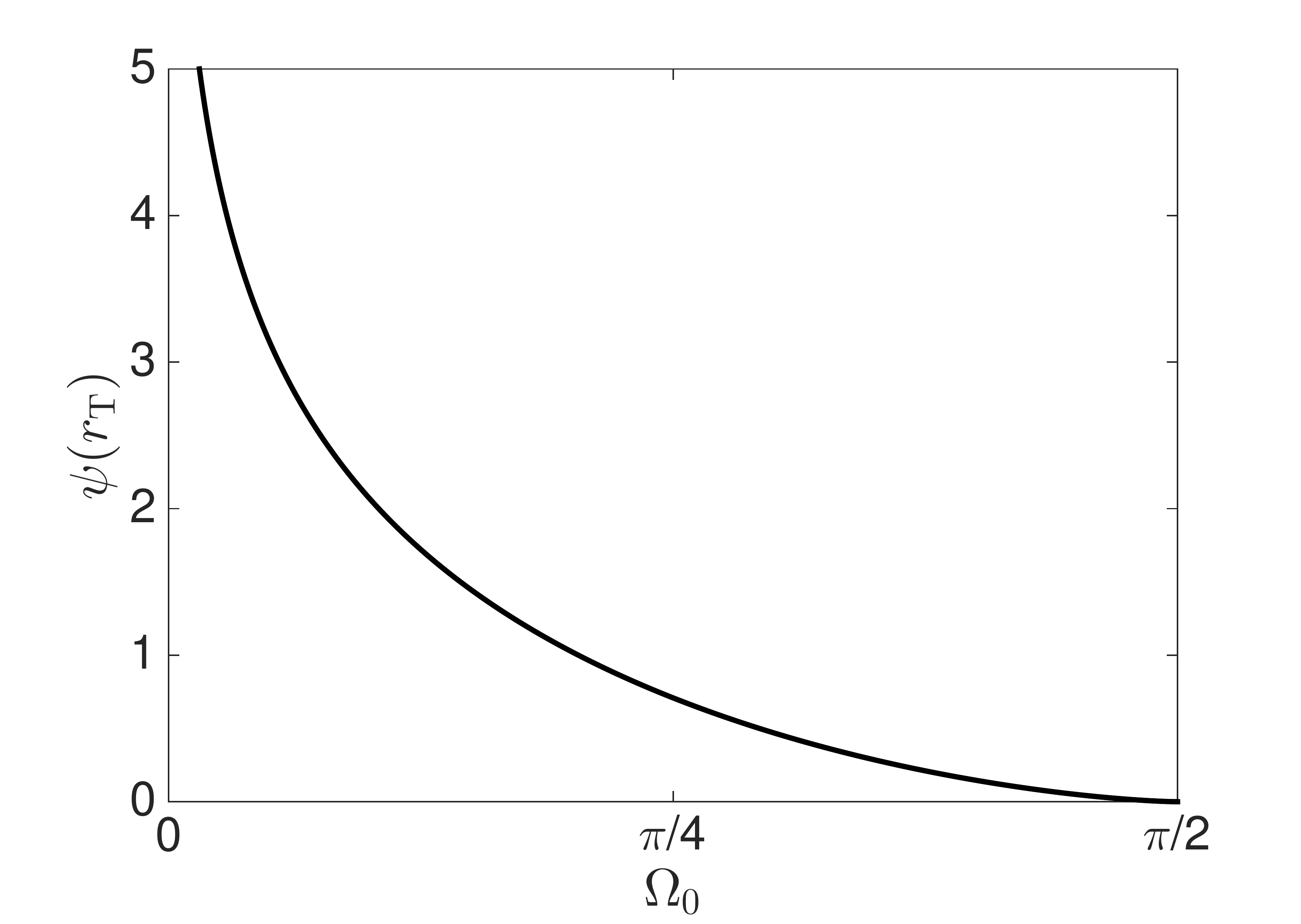}
\caption{The factor $\psi(r_{\text{T}})$ in \eqref{Omi_final} determining the exponential largeness of $\mathcal{Q}$, here shown as a function of the Rayleigh--Bloch frequency $\Omega_0=\pi/(2r_{\text{T}})$}.
\label{fig:psi}
\end{figure}

Formula \eqref{Omi_final} is the main result of this paper. It provides the radiation damping of a structured-ring resonator, which depends on inclusion shape through $\Omega_0$ (or, alternatively, $r_{\text{T}}$) and $\Phi_0$, and also on the mode number $m$. Crucially, the exponential order of magnitude is determined by the function $\psi(r_{\text{T}})$, which is plotted in Fig.~\ref{fig:psi} as a function of $\Omega_0$. From this plot we can directly infer the increase in order of magnitude of the Q-factor as the standing-wave Rayleigh--Bloch frequency $\Omega_0$ is lowered from $\pi/2$ by an appropriate design of the inclusion shape. Conversely, in the limit $\Omega_0\to\pi/2$, i.e. $r_{\text{T}}\to1$, we find from \eqref{psi and h} that $\psi_{\text{T}}$ attenuates like $(r_{\text{T}}-1)^{3/2}$, hinting to the breakdown of the exponential radiation scaling discussed in section \ref{sec:delta}.

\section{Radiation field}\label{sec:pattern}
We have established that in the asymptotic limit $\epsilon\to0$, waves with exponentially small amplitudes are emitted from the cyclic system. Let us consider the radiation field in the whole exterior domain $r>r_\text{T}$, $\phi\in[0,2\pi]$, which in terms of the original polar co-ordinates reads as 
\begin{equation}\begin{split}\label{outer_spiral}
U\sim\frac{1}{(r^2-r_\text{T}^2)^{1/4}}\Big(&D_+e^{i\left\{\varphi(r)+\epsilon mp(r)+\left(\frac{\pi}{2}-\epsilon m\right)\phi\right\}/\epsilon}\\+&D_-e^{i\left\{\varphi(r)-\epsilon mp(r)+\left(-\frac{\pi}{2}-\epsilon m\right)\phi\right\}/\epsilon}\Big).
\end{split}\end{equation}
Note that the exponents in (\ref{outer_spiral}) differ from those in (\ref{ue2}) by a multiplicative phase term $\exp(-im\phi)$. This term represents a relative $O(\epsilon)$ perturbation to the asymptotic solution in the outer-region cell problem, which nevertheless contributes through a cumulative effect at leading order when extended to the full plane.  This term, with which the extended solution appropriately satisfies Bloch's theorem, is consistent with the `secular' part of the solution to \eqref{WKB_e} that is forced by the perturbed Bloch conditions \eqref{anti_m}.

\begin{figure}[H]
  \centering
  \setlength{\unitlength}{\textwidth}
\includegraphics[width=.5\linewidth]{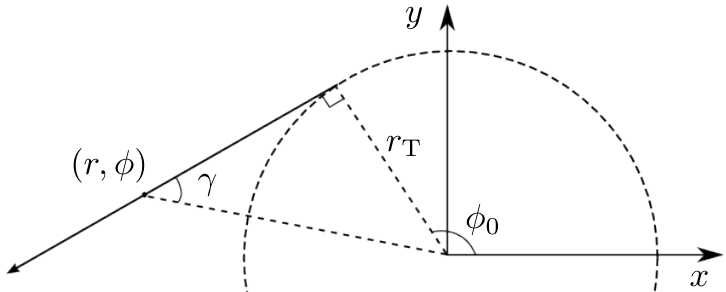}
\caption{Ray direction for a wave described by first term in (\ref{outer_spiral}), given by equation (\ref{line}) with the `+' sign. Here $\gamma=\pi/2+\phi_0-\phi$.}
\label{fig:rt_geom}
\end{figure}

The expression (\ref{outer_spiral}) describes the superposition of two waves. To find the directions in which they propagate, for each term we consider the path of steepest descent, along which
\begin{equation}\label{steep}
\frac{\mathrm{d}r}{\mathrm{d}t}\hat{\mathbf{r}}+r\frac{\mathrm{d}\phi}{\mathrm{d}t}\hat{\bm{\phi}}=-f(t)\nabla\left\{\varphi(r)\pm\epsilon mp(r)+\left(\pm\frac{\pi}{2}\phi-\epsilon m\right)\phi\right\}
\end{equation}
for some parameter $t$ and an unknown function $f$. This leads to a pair of ordinary differential equations that determine the directions of the rays, given by
\begin{equation}\label{ode_polar}
\frac{\mathrm{d}r}{\mathrm{d}\phi}\sim\frac{\Omega_0r\left(r^2/r_\text{T}^2-1\right)^{1/2}\pm\epsilon m r_\text{T}\left(1-r_\text{T}^2/r^2\right)^{-1/2}}{\left(\pm\frac{\pi}{2}-\epsilon m\right)}.
\end{equation}
Let us first consider the case that $m=0$, in which case (\ref{ode_polar}) reduces to the following pair of simple equations
\begin{equation}\label{ode_leading}
\frac{\mathrm{d}r}{\mathrm{d}\phi}\sim \pm\frac{r}{r_\text{T}}\sqrt{r^2-r_\text{T}^2},
\end{equation}
whose solutions are given by
\begin{equation}\label{line}
\phi-\phi_0\sim\pm\left[\frac{\pi}{2}-\operatorname{arccot}\left(\frac{\sqrt{r^2-r_\text{T}^2}}{r_\text{T}}\right)\right].
\end{equation}
We observe from figure \ref{fig:rt_geom} that these equations describe half-lines that are tangent to the circle $r=r_\text{T}$, starting from the point $(r,\phi)=(r_\text{T},\phi_0)$.

\begin{figure}[ht]
  \centering
  \setlength{\unitlength}{\textwidth}
\includegraphics[width=.5\linewidth]{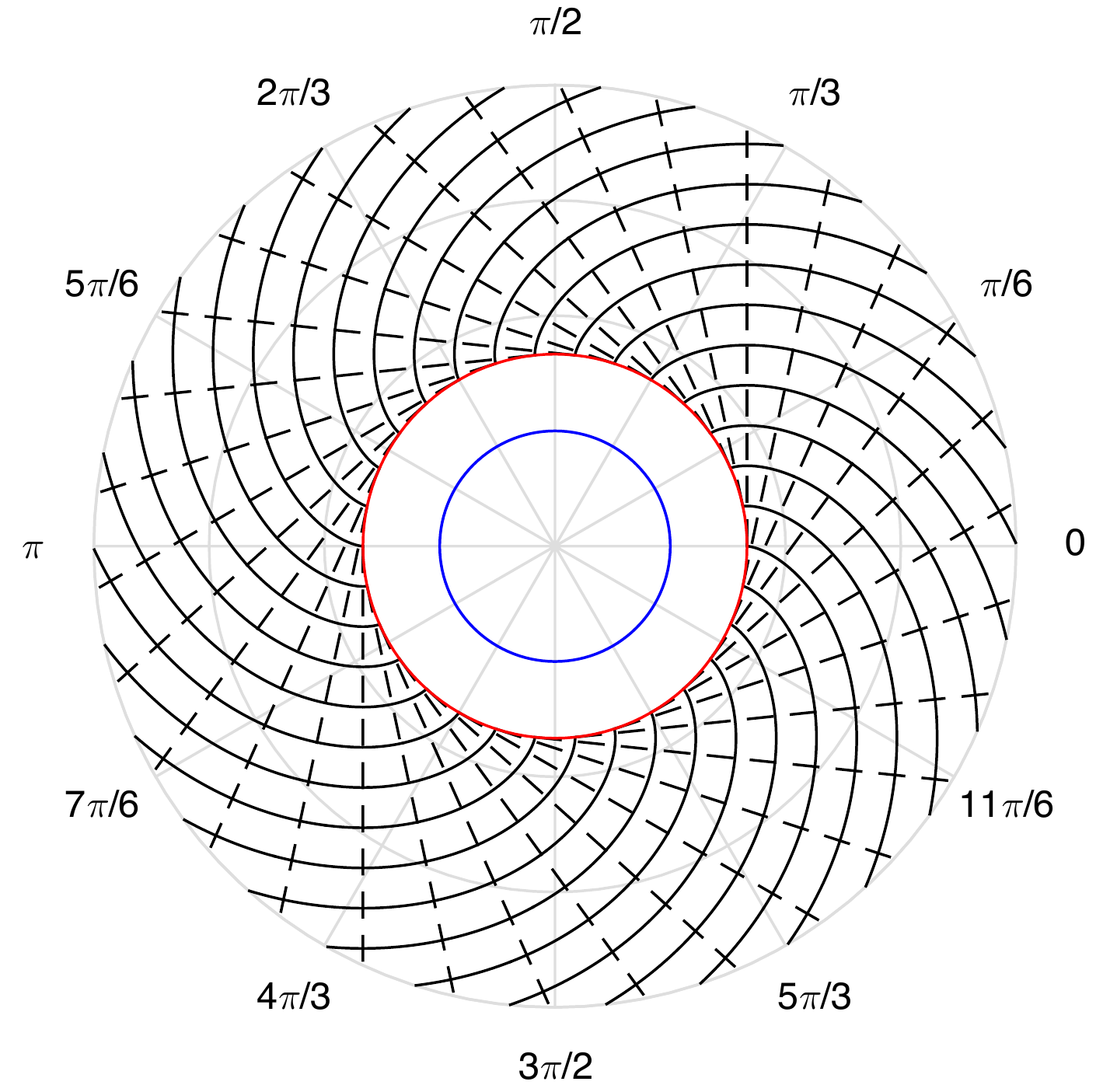}
\caption{Radiation field associated with the first term in (\ref{outer_spiral}) for a ring of slit-like inclusions of length $1.2\epsilon$. The blue circle is $r=1$, the red circle is $r=r_\text{T}$, the dashed black lines are rays given by (\ref{line}), and the solid black curves are wavefronts given by (\ref{archimedes}).}
\label{fig:sketch_spiral}
\end{figure}

For $m\neq0$, it is straightforward to check that expanding (\ref{ode_polar}) to $O(\epsilon)$ leads to the same equation as making the substitution $r_\text{T}\to r_\text{T}\mp\epsilon m/\Omega_0$ in (\ref{ode_leading}) and expanding. To this order, (\ref{ode_polar}) is therefore identical to the equation satisfied by tangent half-lines to the circle $r=r_\text{T}\mp\epsilon m/\Omega_0$, which is the shifted turning point we found in section \ref{ssssec:connection}.
For $m=0$, the wavefronts associated with these rays are given to order $O(\epsilon)$  by
\begin{equation}\label{archimedes}
\phi-\phi_0\sim\mp\frac{2}{\pi}\varphi(r),
\end{equation}
and the adjustment for $m\neq0$ is equivalent to making the substitution $r_\text{T}\to r_\text{T}\mp\epsilon m/\Omega_0$ in the definition of $\varphi(r)$. The resulting field associated with each term in (\ref{outer_spiral}) has a spiral pattern as shown in figure \ref{fig:sketch_spiral} (as $r\to\infty$, $\varphi(r)\sim\Omega_0 r$ so asymptotically (\ref{archimedes}) describes a set of Archimedes spirals). 
Patterns like those described above have been observed in the study of sound fields induced by rotating propellors\cite{prentice_92,chapman_93}, in which case the turning-point ring is referred to as the sonic radius. 

\begin{figure}[ht]
  \centering
  \setlength{\unitlength}{\textwidth}
\includegraphics[width=\linewidth]{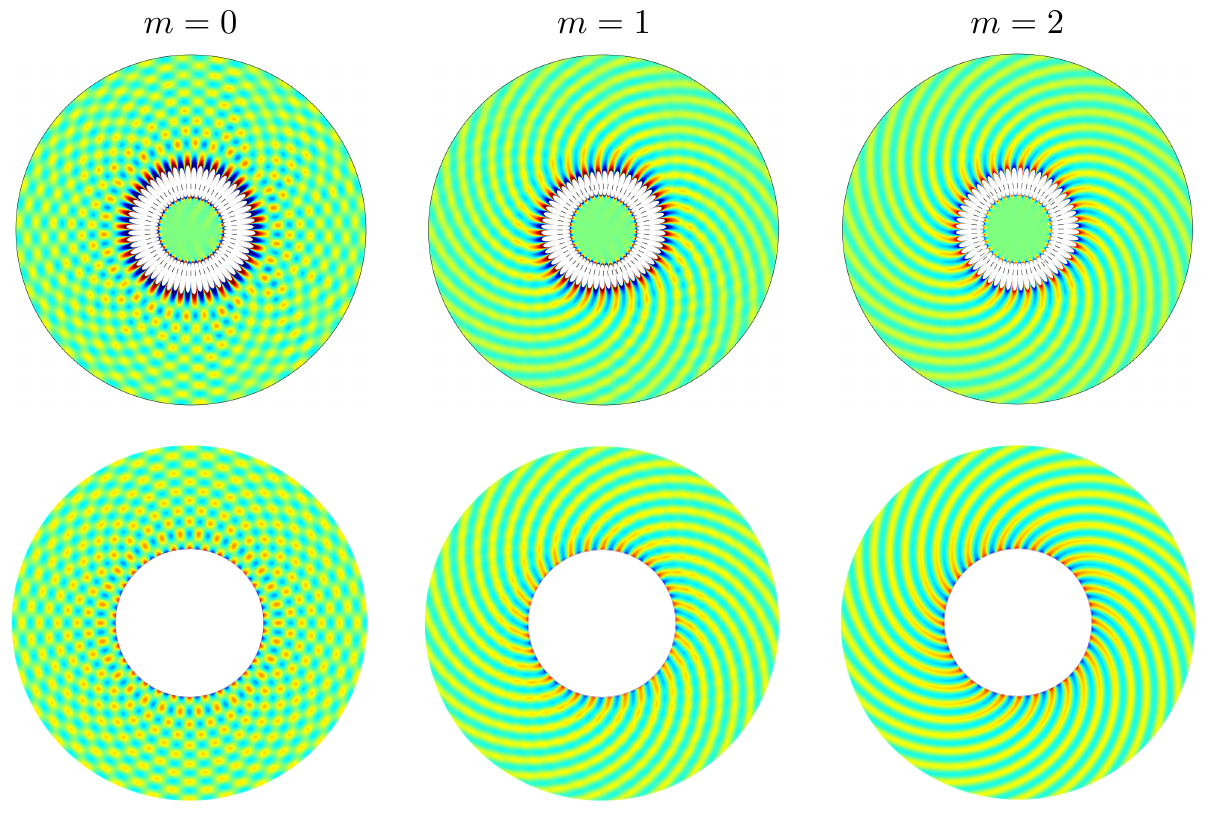}
\caption{Real part of $u(\mathbf{x})$ in the case of 60 slit-like inclusions of length $1.2\epsilon$. The top line shows full finite-element simulations corresponding to the quasi-modes shown in figure \ref{fig:wbm_trio}, but where the colour scale has been saturated, and the bottom line shows the corresponding radiation fields for $r>r_\text{T}$, calculated using the asymptotic formula (\ref{outer_spiral}). In each case the colour scale is linear.}
\label{fig:spirals_1}
\end{figure}

The full radiation field is a superposition of the two terms in (\ref{outer_spiral}), whose opposing ray directions result in wavefronts that bend and precess in opposite directions to each other. In the anti-periodic case, the two terms have equal weighting as $|D_+|=|D_-|$, resulting in an interference pattern like the one seen in the left hand column of figure \ref{fig:spirals_1}. For $m\neq0$, the factor $e^{\mp mh(r_\text{T})}$ in (\ref{D}) causes one term to dominate over the other, so one spiral is clearly distinguishable, as seen in the right two columns in figure \ref{fig:spirals_1}.

\section{Intermediate asymptotics of algebraic radiation loss}\label{sec:delta}
The preceding analysis hinges upon the assumption that the Rayleigh--Bloch frequency $\Omega_0$ is sufficiently far below $\pi/2$ that the turning point ring is separated from the inner region containing the inclusions by an $O(1)$ distance. While a detailed analysis of the case where $\pi/2-\Omega_0$ is small is outside the scope of this paper, we make the following comments. For fixed $\pi/2-\Omega_0$, and $\epsilon=\pi/N$ sufficiently small, we expect our asymptotic analysis to continue to hold to leading order. As $N$ decreases, however, we expect our asymptotic analysis to break down as the radius $\approx r_{\text{T}}$ of the turning-point ring shrinks towards the ring of inclusions; as the tunnelling distance vanishes, we anticipate a transition via a series of intermediate asymptotic limits from exponential to algebraic radiation damping. Recall in particular that in section \ref{sec:damping} we found that the exponential scaling of \eqref{Omi_final} breaks down when $(\pi/2-\Omega_0)$ and $(r_{\text{T}}-1)$ become comparable to $\epsilon^{2/3}$, which corresponds to the limit in which the $O(\epsilon^{2/3})$ transition region overlaps with the inner region.

\begin{figure}
  \centering
  \setlength{\unitlength}{\textwidth}
\includegraphics[width=.9\linewidth]{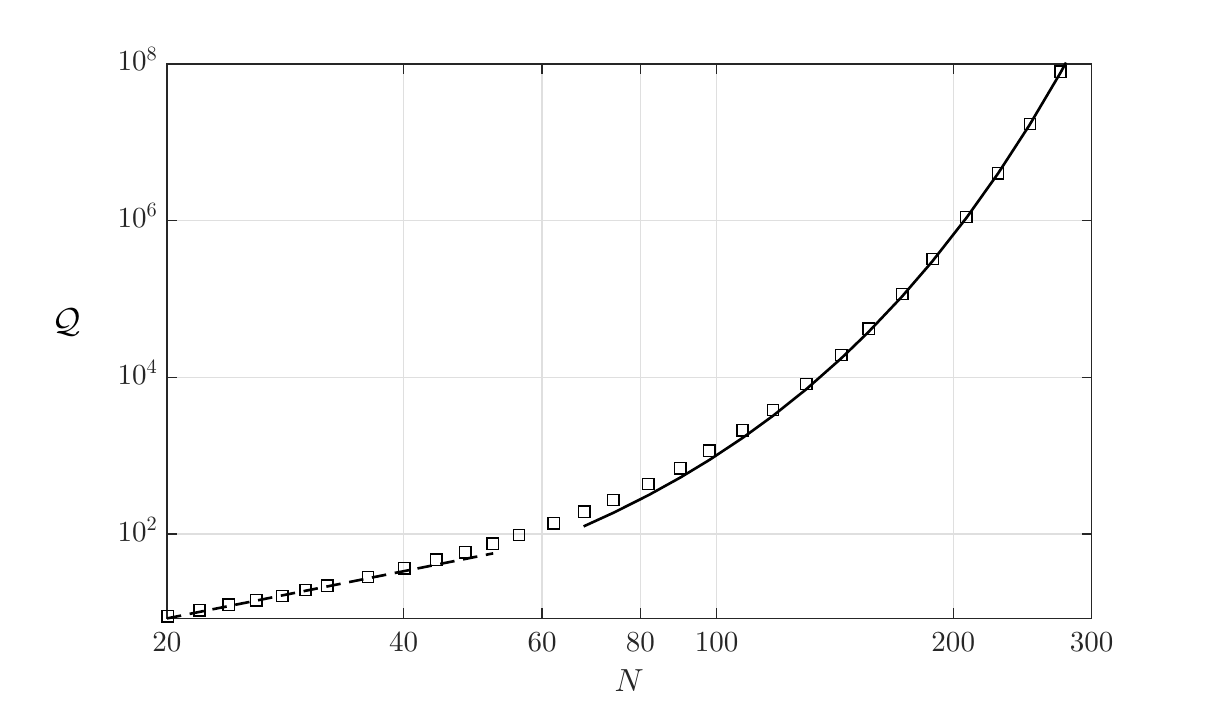}
\caption{Dependence of $\mathcal{Q}$ on $N$ for a ring of circular inclusions of radius $0.8\epsilon$ for $m=0$. The squares are from a 12-term multipole expansion treatment, the solid line is from the asymptotic formula (\ref{Omi_final}), and the dashed line is an $N^2$ fit.}
\label{fig:both_ends}
\end{figure}

To demonstrate this, in figure \ref{fig:both_ends} we calculate $\mathcal{Q}$ for rings of circular holes of radius $0.8\epsilon$, for varying values of $N$. From the associated Rayleigh--Bloch eigenvalue problem, we calculate $\Omega_0=1.321$, which gives $r_\text{T}-1=0.189$. As expected, for very large values of $N$, our asymptotic method captures the correct behaviour, but as $N$ is decreased (and $\epsilon$ increased), the curve transitions to having algebraic dependence, with $\mathcal{Q}=O(N^2)$ for $N\lesssim40$; note that $\epsilon^{2/3}=(\pi/40)^{2/3}\approx 0.183$ is comparable to $r_{\text{T}}-1$, in agreement with our above estimate for the breakdown of the exponential regime. As a final comment, we expect the transition from exponential to algebraic damping to occur sooner for mode numbers with $m\neq0$, as the turning point for one of the terms moves closer to the ring of inclusions; this is consistent with what is seen in figure \ref{fig:Omi}.

\section{Concluding remarks}\label{sec:conclude}
We have analysed the radiating quasi-normal modes of structured-ring resonators in the limit of a large number of inclusions. The asymptotics lead to a deeper understanding of the physical origin of the observed phenomena as well as generating the asymptotic formula \eqref{Omi_final} for the exponentially large Q-factor. The latter is given entirely in terms of properties of the standing-wave Rayleigh--Bloch modes supported by the corresponding linear array, along with the number of inclusions $N$ and the prescribed cyclic mode number $m$; for the design of structured-ring resonators this allows one to directly harness previous work on Rayleigh--Bloch waves. In particular, our formula explicitly shows how the exponential asymptotic order of the loss is determined by the separation between the Rayleigh--Bloch frequency and the light (sound) line. 

We have restricted our attention to moderate cyclic mode numbers $m$, which ensures the resonance frequency is close to the Rayleigh--Bloch standing-wave frequency, and inclusion shapes for which the latter frequency is not too close to the light-line crossing at $\Omega=\pi/2$.
As $\Omega_0\to\pi/2$, or with increasing $m$, the resonance frequency approaches the light line, and at least one of the turning-point rings shrinks towards the ring of inclusions. In the former limit, we demonstrated a transition with increasing $N$ between algebraic and exponential radiation damping. For large $m$, there can be a turning-point ring for $r<1$, with a wave bouncing back and forth between the inclusion array and the internal turning point ring; this has been observed in Ref.~\cite{maling_16_wbm}.

Lastly we emphasise that the regime of interest here is implicitly that of high frequency, or equivalently of short wavelengths commensurate with the array spacing, wherein the existing modes are locally guided by the ring of inclusions. Our analysis is therefore  complementary to the low-frequency homogenisation approaches recently used for the Faraday cage \cite{chapman_15,Hewett_16}, where the energy of the resonances is stored over the interior of the ring as opposed to being localised to its circumference.

\section{Funding}\label{funding}
This work was funded by ESPRC UK Programme Grant EP/L024926/1.

\bibliographystyle{unsrt}

\bibliography{RB_WKB}

\begin{thebibliography}{10}

\bibitem{vahala_03}
K.~J. Vahala.
\newblock Optical microcavities.
\newblock {\em Nature}, 424:839--846, 2003.

\bibitem{whisper_rev}
A.~B. Matsko, A.~A. Savchenkov, D.~Strekalov, V.~S. Ilchenko, and L.~Maleki.
\newblock Review of applications of whispering-gallery mode resonators in
  photonics and nonlinear optics.
\newblock {\em IPN Progress Report}, 2005.

\bibitem{vollmer_08}
F.~Vollmer and S.~Arnold.
\newblock Whispering-gallery-mode biosensing: label-free detection down to
  single molecules.
\newblock {\em Nature Methods}, 5:591--596, 2008.

\bibitem{Soria_11}
S.~Soria, S.~Berneschi, M.~Brenci, F.~Cosi, G.~N. Conti, S.~Pelli, and G.~C.
  Righini.
\newblock Optical microspherical resonators for biomedical sensing.
\newblock {\em Sensors}, 11:785--805, 2011.

\bibitem{dixit_01}
S.~K. Dixit.
\newblock {\em Filtering resonators}.
\newblock Nova science publishers, 2001.

\bibitem{hall_89}
D.~Hall and P.~Jackson.
\newblock {\em The physics and technology of laser resonators}.
\newblock Taylor and Francis, 1989.

\bibitem{kudryashov_99}
A.~Kudryashov and H.~Weber.
\newblock {\em Laser resonators: novel design and development}.
\newblock SPIE press, 1999.

\bibitem{bykov_95}
V.~Bykov and O~Silichev.
\newblock {\em Laser resonators}.
\newblock Cambridge Intl. Science Publ., 1995.

\bibitem{hodgson_05}
N.~Hodgson and H.~Weber.
\newblock {\em Laser resonators and beam propagation, 2nd edition}.
\newblock Springer, 2005.

\bibitem{ilchenko_04}
V.~S. Ilchenko, A.~A. Savchenko, A.~B. Matsko, and L.~Maleki.
\newblock Nonlinear optics and crystalline whispering gallery mode cavities.
\newblock {\em Phys. Rev. Lett.}, 92, 2004.

\bibitem{hofer_10}
J.~Hofer, A.~Schliesser, and T.~J. Kippenberg.
\newblock Cavity optomechanics with ultrahigh-{Q} crystalline microresonators.
\newblock {\em Phys. Rev. A}, 82, 2010.

\bibitem{rabus_07}
D.~G. Rabus.
\newblock {\em Integrated ring resonators: the compendium}.
\newblock Springer, 2007.

\bibitem{righini11a}
G.~C. Righini, Y.~Dumeige, P.~F\'eron, M.~Ferrari, G.~{Nunzi Conti}, D.~Ristic,
  and S.~Soria.
\newblock Whispering gallery mode microresonators: {F}undamentals and
  applications.
\newblock {\em Rivista Del Nuovo Cimento}, 34:435--488, 2011.

\bibitem{melloni01a}
A.~Melloni.
\newblock Synthesis of a parallel-coupled ring-resonator filter.
\newblock {\em Opt. Lett.}, 26(12):917--919, Jun 2001.

\bibitem{chin98a}
M.~K. Chin and S.~T. Ho.
\newblock Design and modeling of waveguide-coupled single-mode microring
  resonators.
\newblock {\em Journal of Lightwave Technology}, 16(8):1433, 1998.

\bibitem{Heiblum:75}
M.~Heiblum and J.~H. Harris.
\newblock Analysis of curved optical waveguides by conformal transformation.
\newblock {\em IEEE J. Quantum Electron.}, 11:75--83, 1975.

\bibitem{snyder83a}
A.~W. Snyder and J.~D. Love.
\newblock {\em Optical waveguide theory}.
\newblock Chapman and Hall Ltd, 1983.

\bibitem{Hurd_54}
R.~A. Hurd.
\newblock The propagation of an electromagnetic wave along an infinite
  corrugated surface.
\newblock {\em Can. J. Phys.}, 32:727--734, 1954.

\bibitem{Sengupta_59}
D.~Sengupta.
\newblock On the phase velocity of wave propagation along an infinite yagi
  structure.
\newblock {\em IRE Trans. Antennas Propag.}, 7:234--239, 1959.

\bibitem{Evans_93}
D.~V. Evans and C.~M. Linton.
\newblock Edge waves along periodic coastlines.
\newblock {\em Q. J. Appl. Math}, 46:643--656, 1993.

\bibitem{pendry04a}
J.~B. Pendry, L.~Martin-Moreno, and F.~J. Garcia-Vidal.
\newblock Mimicking surface plasmons with structured surfaces.
\newblock {\em Science}, 305:847--848, 2004.

\bibitem{every08a}
A.~G. Every.
\newblock Guided elastic waves at a periodic array of thin coplanar cavities in
  a solid.
\newblock {\em Phys. Rev. B}, 78:174104, 2008.

\bibitem{colquitt14a}
D.~J. Colquitt, R.~V. Craster, T.~Antonakakis, and S.~Guennaeu.
\newblock Rayleigh-{B}loch waves along elastic diffraction gratings.
\newblock {\em Proc. R. Soc. Lond. A}, 471:20140465, 2015.

\bibitem{bonnet94a}
A.~S. {Bonnet-Bendhia} and F.~Starling.
\newblock Guided waves by electromagnetic gratings and nonuniqueness examples
  for the diffraction problem.
\newblock {\em Math. Meth. Appl. Sci.}, 17:305--338, 1994.

\bibitem{Linton_02}
C.~M. Linton and M.~McIver.
\newblock The existence of {R}ayleigh-{B}loch surface waves.
\newblock {\em J. Fluid Mech.}, 470:85--90, 2002.

\bibitem{porter99a}
R.~Porter and D.~V. Evans.
\newblock Rayleigh-{B}loch surface waves along periodic gratings and their
  connection with trapped modes in waveguides.
\newblock {\em J. Fluid Mech.}, 386:233--258, 1999.

\bibitem{thompson08a}
I.~Thompson and Porter R.
\newblock A new approximation method for scattering by long finite arrays.
\newblock {\em Q. J. Mech. Appl. Math.}, 61:234--239, 2008.

\bibitem{pors_12}
A.~Pors, E.~Moreno, L.~Martin-Moreno, J.~B. Pendry, and F.~J. Garcia-Vidal.
\newblock Localized spoof plasmons arise while texturing closed surfaces.
\newblock {\em Phys. Rev. Lett.}, 108:223905, May 2012.

\bibitem{paloma}
P.~A. Huidubro, S.~Xiaopeng, J.~Cuerda, E.~Moreno, L.~Martin-Moreno, F.~J.
  Garcia-Vidal, T.~J. Cui, and J.~B. Pendry.
\newblock Magnetic localized surface plasmons.
\newblock {\em Phys. Rev. X.}, 4, 2014.

\bibitem{linton02a}
C.~M. Linton and M.~Mc{I}ver.
\newblock The existence of {R}ayleig{h-B}loch surface waves.
\newblock {\em J. Fluid Mech.}, 470:85--90, 2002.

\bibitem{comsol}
COMSOL ltd.
\newblock Comsol multiphysics 5.0, 2014.

\bibitem{maling_16_wbm}
B.~Maling and R.~V. Craster.
\newblock Whispering {B}loch modes.
\newblock {\em Proc. R. Soc. A}, 472:20160103, 2016.

\bibitem{bender}
C.~M. Bender and S.~A. Orszag.
\newblock {\em Advanced mathematical methods for scientists and engineers}.
\newblock McGraw-Hill, New York, 1978.

\bibitem{hinch}
E.~J. Hinch.
\newblock {\em Perturbation methods}.
\newblock Cambridge University Press, 1991.

\bibitem{gridin03b}
D.~Gridin and R.~V. Craster.
\newblock Quasi-modes of a weakly curved waveguide.
\newblock {\em Proc. R. Soc. Lond. A}, 459:2909--2931, 2003.

\bibitem{ching98a}
E.~S.~C. Ching, P.~T. Leung, A.~{Maassen van den Brink}, W.~M. Suen, S.~S.
  Tong, and K.~Young.
\newblock Quasinormal-mode expansion for waves in open systems.
\newblock {\em Rev. Mod. Phys.}, 1998.

\bibitem{Collin}
R.~E. Collin.
\newblock {\em Foundations of microwave engineering}.
\newblock McGraw Hill Book Company, Inc., New York, 1966.

\bibitem{Maier:07}
S.~A. Maier.
\newblock {\em Plasmonics: fundamentals and applications}.
\newblock Springer Science \& Business Media, 2007.

\bibitem{Hewett_16}
D.~P. Hewett and I.~J. Hewitt.
\newblock Homogenized boundary conditions and resonance effects in {F}araday
  cages.
\newblock {\em Proc. R. Soc. A}, 472(2189), 2016.

\bibitem{prentice_92}
P.~R. Prentice.
\newblock The acoustic ring source and its application to propeller acoustics.
\newblock {\em Proc. R. Soc. A}, 437, 1992.

\bibitem{chapman_93}
C.~J. Chapman.
\newblock The structure of rotating sound fields.
\newblock {\em Proc. R. Soc. A}, 440, 1993.

\bibitem{chapman_15}
S.~J. Chapman, D.P. Hewett, and L.~N. Trefethen.
\newblock Mathematics of the {F}araday cage.
\newblock {\em SIAM Rev.}, 57:398--417, 2015.

\end{thebibliography}

\end{document}